\documentclass[12pt,preprint]{aastex}

\def\kms  {km~s$^{-1}$}
\def\masy {mas~y$^{-1}$}

\def\etal {et al.~}
\def\eg   {e.g.~}
\def\ie   {i.e.~}
\def\hho  {H$_2$O}
\def\meth {CH$_3$OH}
\def\Ssrc {S~252}
\def\Gsrc {G232.6+1.0}
\def\Vlsr {\ifmmode {V_{\rm LSR}} \else {$V_{\rm LSR}$} \fi}
\def\Ro   {\ifmmode {R_0} \else {$R_0$} \fi}
\def\To   {\ifmmode {\Theta_0} \else {$\Theta_0$} \fi}
\def\uvdata{{\it (u,v)}-data}

\slugcomment{Version: 2008 Oct 27}
\shorttitle{Trigonometric Parallaxes of \Ssrc\ \& \Gsrc} 
\shortauthors{Reid \etal}

\begin{document}

\title{Trigonometric Parallaxes of Massive Star Forming Regions: I.
          \\ \Ssrc\ \& \Gsrc}

\author{M. J. Reid\altaffilmark{1}, K. M. Menten\altaffilmark{2}, 
        A. Brunthaler\altaffilmark{2}, X. W. Zheng\altaffilmark{3}, 
        L. Moscadelli\altaffilmark{4} and Y. Xu\altaffilmark{2,5}}

\altaffiltext{1}{Harvard-Smithsonian Center for
   Astrophysics, 60 Garden Street, Cambridge, MA 02138, USA}
\altaffiltext{2}{Max-Planck-Institut f\"ur Radioastronomie, 
   Auf dem H\"ugel 69, 53121 Bonn, Germany}
\altaffiltext{3}{Department of Astronomy, Nanjing University
   Nanjing 210093, China} 
\altaffiltext{4}{INAF, Osservatorio Astrofisico di Arcetri, 
   Largo E. Fermi 5, 50125 Firenze, Italy}
\altaffiltext{5}{Purple Mountain Observatory, Chinese Academy of
   Sciences, Nanjing 210008, China}

\begin{abstract}
We are conducting a large program with the NRAO 
Very Long Baseline Array (VLBA) to measure trigonometric parallaxes 
of massive star-forming regions across the Milky Way. 
Here we report measurement of the parallax and proper motion of
methanol masers in \Ssrc\ and \Gsrc. 
The parallax of \Ssrc\ is $0.476\pm0.006$~mas 
($2.10^{+0.027}_{-0.026}$~kpc), placing it in the Perseus spiral arm.
The parallax of \Gsrc\ is $0.596 \pm 0.035$~mas 
($1.68^{+0.11}_{-0.09}$~kpc),
placing it between the Carina-Sagittarius and Perseus arms, possibly in 
a Local (Orion) spur of the Carina-Sagittarius arm.
For both sources, kinematic distances are significantly
greater than their parallax distances.  Our parallaxes and proper motions 
yield full space motions accurate to $\approx 1$~\kms.
Both sources orbit the Galaxy $\sim13$~\kms\ slower than circular
rotation. 
\end{abstract}


\section{Introduction}

An image of the Milky Way, taken by an observer in another galaxy would 
probably reveal a spiral structure dotted with many bright
HII regions.  The Milky Way is believed to be spiral galaxy, and a best
``educated guess'' is that it is a barred Sb to Sc galaxy
\citep{Blitz:83,Gerhard:02}.
However, since we are inside the Milky Way, it has proved very difficult 
to properly characterize its structure \citep{Bash:81,Burton:88}.

Originally, studies of HI emission offered the opportunity to map the
structure of the Milky Way \citep{Oort:58}.
HI emission on longitude-velocity plots clearly demonstrated that there
were some coherent, large-scale structures, which were probably spiral
arms in the Milky Way.  However, determining accurate distances to HI
clouds proved problematic, and this made the task of turning
longitude-velocity data into a true plan-view of the Milky Way very 
uncertain \citep{Burton:88}.
Later, millimeter-wave observations of molecules, such as CO, also revealed
coherent, large-scale structures with higher contrast than seen in HI
\citep{Dame:01}.
But, again, uncertain distances to molecular clouds precluded making a true
map of the Milky Way with sufficient accuracy map the spiral structure 
of the Milky Way.

\citet{Georgelin:76} published 
a ``plan-view'' model of the spiral structure of the Milky Way.  
This approach involved combining optical
observations of young stars and radio data of HI cloud and HII region
emissions.  Luminosity distances to nearby stars were used where available
and kinematic distances elsewhere, mostly for more distant HII regions.
More recently, \citet{Taylor:93} have modeled pulsar dispersion measures 
to refine the Georgelin \& Georgelin model.  While subject to very significant 
uncertainties from kinematic distances, the Georgelin \& Georgelin model has 
remained the basis for the ``standard'' model of the spiral structure of the 
Milky Way for over 30 years.  

The primary goal of our project is to reinforce the foundations of models
of the spiral structure of the Milky Way by measuring distances directly
to regions of massive star formation across large portions of the Milky Way.
This paper is the first (Paper I) in a series, including Papers II through V 
\citep{Moscadelli:09,Xu:09,Zhang:09,Brunthaler:09} published in this volume.
We are using the NRAO 
\footnote{The National Radio Astronomy Observatory is a facility of the 
National Science Foundation operated under cooperative agreement by 
Associated Universities, Inc.}
Very Long Baseline Array (VLBA) to determine trigonometric parallaxes 
of strong methanol maser sources, which are associated with regions of 
massive star formation and their attendant HII regions.  
Sampling spiral arms roughly every kpc should determine the true locations 
of arms, and allow us to use other databases to ``interpolate'' between the 
star forming regions measured with masers.
With accurate distances to some of the largest star forming regions,
we should be able to verify the existence and determine the locations
of the postulated Perseus, Carina--Sagittarius, Crux--Scutum, and 
Norma spiral arms.
Ultimately, we hope to extend these measurements with a larger sample,
including a similar study in the southern hemisphere, and produce a
map of the 3-dimensional structure of bright material
associated with massive young stars that defines spiral structure.

In addition to distances, the observations used to determine trigonometric
parallaxes yield excellent measurements of secular proper motions, with
accuracies of $\approx 1$~\kms\ \citep{Xu:06a}.
Combining radial velocity measurements with proper motions (and distances)
yields full 3-dimensional velocities, relative to the motion of the Sun.
Thus, through this project and other VLBI efforts, notably 
the Japanese project VERA \citep{Kobayashi:05},
we hope to determine the full kinematics of massive star forming regions
in the Milky Way, which will accurately define the rotation curve of the
Milky Way and, in turn, its enclosed mass as a function of Galactocentric
radius.  Finally, we should be able to show how material in spiral arms
actually moves, to characterize kinematic anomalies (such as in the 
Perseus Arm) and, hopefully, to understand why these anomalies occur.

\section {Radio Wavelength Parallax Essentials}

\subsection{Methanol Masers}

Methanol (\meth) masers are excellent astrometric targets for parallax
measurements.  Class II methanol masers
(\eg 6.7 and 12~GHz transitions) are widespread and associated with
newly formed stars in regions of high-mass star formation.  
The masers generally are compact ($\sim1$~mas), slow moving, and vary 
slowly, which minimizes the possibility of brightness variations mimicking 
position shifts.  While the 6.7~GHz masing transition is generally stronger 
than the 12~GHz transition, there are dozens of 12~GHz sources with peak
flux densities $>5$~Jy, which is sufficient to serve as a phase-reference
for the VLBA.   Also, the 12~GHz transition is at a high enough frequency 
to offer minimal sensitivity to unmodeled ionospheric fluctuations and
to minimize interstellar scatter broadening.  However, once all 12~GHz
masers have been measured, the 6.7~GHz methanol masers will be attractive
targets.  

We note that 22~GHz \hho\ masers are also good astrometric targets; 
they are compact, strong, widespread, and the high frequency of 
the transition minimizes ionospheric fluctuations and interstellar scattering 
problems. However, \hho\ masers can be variable on time scales as short 
as weeks to months and, since parallax observations are best made over a 
timespan of 1~yr, this can be problematic.  Water masers are generally
associated with high-velocity outflows from young stellar objects.
Since, for well-planned measurements, parallax and proper motion
are essentially uncorrelated, parallax estimates
should not be affected by the magnitude of the proper motion.  However,
one of the most astrophyscially interesting by-products of maser 
astrometry is the determination of the full space-motions of the
associated young stellar objects.   If there are fast internal motions,
then one needs to understand details of the outflows in order to transform
from measured maser motions to a frame tied to the central stellar object.
The accuracy of this transformation is typically better than a few \kms\
for slowly expanding methanol masers \citep{Xu:06a}, but can be of order
10 \kms\ for the rapidly expanding \hho\ masers \citep{Hachisuka:06}.

\subsection{Identifying Background Reference Sources}

Since astrometric accuracy is enhanced by finding background sources 
(ie, QSOs) as close in angle to the maser sources as possible,
we first conducted surveys for background sources near potential
maser targets.   Using the sustainable recording rate of
128 Mbits s$^{-1}$ of the VLBA in 2006 (which is being substantially 
upgraded), we are able to use background sources as weak as a 
few mJy.  We identified candidate background sources by selecting all 
sources in the NRAO VLA Sky Survey (NVSS) that were nearly unresolved 
(at $\approx40''$ resolution) and within $\approx1.5^\circ$
of a maser target.  Typically we found $\approx30$ candidates stronger 
than $\approx20$~mJy for any maser target.
These candidate sources were observed in snap-shot mode 
with the VLA at 8 and 15~GHz to determine a spectral index and further 
test compactness.  Candidates that remained unresolved with these
observations and displayed a synchrotron spectral index are likely to
be extragalactic and have compact cores that can be imaged with VLBA
observations.  Typically we find between 1 and 3 suitable QSOs within about
1.5 degrees of a target methanol maser source \citep{Xu:06b}.
When possible we re-observed with these sources with the VLA
A- or B-configurations to determine absolute positions to better than 
30~mas.  

\subsection{Atmospheric Delay Calibration} \label{atm_cal}

The main source of systematic error for cm-wave phase-referenced 
observations is uncompensated interferometric delays introduced 
by the Earth's atmosphere and ionosphere.
The model used in the VLBA correlator is a seasonally-averaged
calculation that does not take into account variations in atmospheric
pressure, total water-vapor content, nor the delay induced by the
ionosphere.  At a frequency of 12~GHz the correlator model will typically
be in error by about 0.2 to 0.4 nsec of delay, corresponding to 
5 to 10~cm of excess path length, for sources at high elevation angles.
Substantial gains in reducing systematic errors owing to atmospheric 
mis-modeling can be obtained by directly measuring and removing these 
effects.  

For the last decade, we have been working to understand better our
sources of systematic error and to find methods of correcting our
data for these errors \citep{Reid:99,Reid:04}
Currently, we use observations of extragalactic sources 
spread over the sky to measure broad-band delays.  For sources with well 
determined positions (better than 1~mas), residual delays measured
by the VLBA will be dominated by the effects of atmospheric mis-modeling.
Typically, we observe $\approx15$ such sources over a range of source
azimuths and elevations in rapid succession over a time span of about
40 min.  

The choice of sources and the observing sequence were
determined by computer code with the following procedure:
We generated a candidate source list starting with the ICRF catalog
\citep{Fey:04} and keeping only the 60 sources that had been observed
$>1000$ times, had coordinate uncertainties $<1$~mas, and had
little structure at 8 GHz (\ie\ structural indexes of 1 or 2). 
Trial schedules were generated by 
randomly skipping though the candidate source list; we required that 
a source was above an elevation of $8^\circ$ at a minimum of 6 VLBA 
stations and that no station missed more than a total of 5 sources. 
For each trial schedule, we generated fake data and 
performed a least-squares fit solving for zenith atmospheric delays
and clock offsets for all stations.   We generally examined 1000 trial 
schedules and chose the one that minimized the largest of the formal 
zenith delay uncertainties among the 10 stations.

We placed these ``geodetic'' blocks before the start, in 
the middle, and after the end of our phase-reference observations, in order 
to monitor slow changes in the total atmospheric delay for each telescope.   
The data were taken in left circular polarization 
with eight 4-MHz bands that spanned 480 MHz of bandwidth between 
12.10 and 12.58 GHz; the bands were spaced in a 
``minimum redundancy configuration'' to uniformly sample, as best as possible, 
all frequency differences.  This was accomplished with bands spaced by 
0, 1, 4, 9, 15, 22, 32 and 34 times 14 MHz.   The data were
correlated, corrected for ionospheric delays using total electron content 
measurements \citep{Walker:00}, and residual multi-band delays 
and fringe rates were determined for all sources.  

\clearpage

\begin{figure}
\epsscale{0.45} 
\plotone{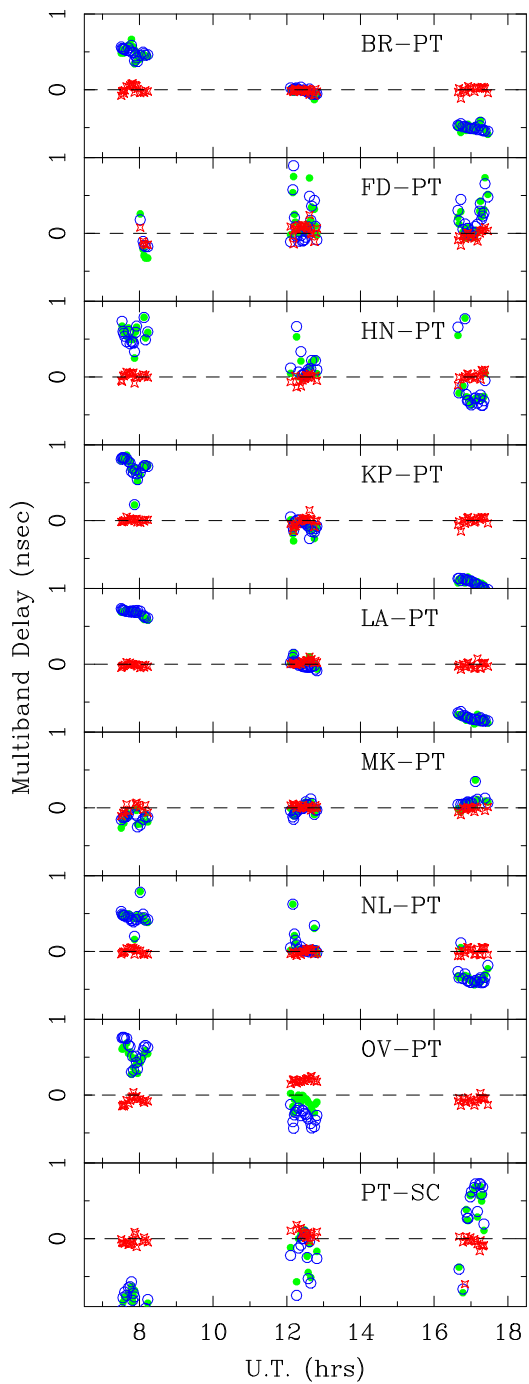} 
\caption{Data (green dots), model (blue circles), and residuals
(red crosses) for multi-band delays from a solution to all three 
geodetic blocks simultaneously.
Only residuals for baselines involving the reference antenna
(PT) are plotted; standard VLBA 2-character antenna names indicate
the interferometer baseline.
For each antenna, we solved for a vertical 
atmospheric delay and its time derivative and, except for the 
reference antenna, a clock and its time derivative. 
NB: 0.1 nsec~$\approx 3$~cm of path delay.
  \label{fit_geodetic}
        }
\end{figure}

\clearpage

The multi-band delays and fringe rates were modeled as owing 
to a vertical atmospheric delay and delay-rate, as well as a clock
offset and clock drift rate, at each antenna.  Owing to the
short (1~min) integration times, the fringe rates are fairly uncertain 
and the multi-band delays carry most of the weight in the solutions.
In order to protect the fitting process from occasional bad data,
we used an iterative approach that down-weights outliers relative to
the previous solution.  Specifically, after each iteration, the
uncertainty assigned to the each delay or fringe rate datum was
increased by a factor of ${\rm exp}[(r/4)^2]$, 
where $r$ is the post-fit residual divided by the original uncertainty.
This procedure heavily down-weights $>4\sigma$ outliers.
Typically, only a few percent of the input data were significantly
down-weighted.  Sample data, model, and residuals from our first epoch 
observations are shown in Fig.~\ref{fit_geodetic}.

In some cases, the delay residuals indicated 
departures from the linear (in time) model; an example can be seen in
the non-zero mean delay residuals ({\it red crosses}) in 
Fig.~\ref{fit_geodetic} for the interferometer baseline formed by 
VLBA antennas OV and PT.
We believe this is due to unusually large, quasi-random, changes in the 
zenith delay at antenna 8 during this observation.  In order to
improve the calibration, we then solved for zenith delays and
clock offsets for each 40-min block separately.  Calibration of
the phase-referencing data was done by linearly interpolating between
these three solutions.

Using our least-squares fitting program, we estimated zenith atmospheric 
path-delays with accuracies typically $<1.0$~cm ($<0.4\lambda$).
Independent comparisons of our technique with those using Global 
Positioning System (GPS) data and an image-optimization approach
confirm this accuracy \citep{Honma:08}. 
Correction for atmospheric delays results in significant 
improvement in relative position accuracy, compared to 
observations that rely solely on the VLBA correlator model. Finally, 
using more than one background reference source, at different offsets and 
position angles with respect to the target maser source, when
phase-referencing further reduces sensitivity to any residual 
systematic errors from atmospheric mis-modeling.

\subsection{Absolute Positions} \label{abs_pos}

When conducting phase-referenced observations, it is important
that the position of the reference source is well determined
\citep{Beasley:95}.
A position error, $\Delta\theta_{err}$, for the reference source results 
in residual phases shifts that are sinusoidal with a 24 hour period 
and an amplitude of $\sim\Delta\theta_{err}/\theta_f$~turns, 
where $\theta_f$ is the interferometer fringe spacing.  
When applied to a target source, offset from the reference source by 
$\theta_{sep}$, the reference phases can no longer be perfectly 
modeled as a position shift (at the position of the target source).  
This results in a second-order position shift of 
$\sim\Delta\theta_{err}\theta_{sep}$, as well a degradation
of the imaging.  While the second-order position shift, in principle,
can be constant from epoch to epoch and not affect a parallax
measurement, the image degradation can be serious and even preclude
detection of a weak source.   

We attempted to determine the absolute positions 
of our sources to better than $10$~mas accuracy.  
The 12 GHz methanol masers for which we sought to measure parallaxes 
contain some bright, compact spots that serve as good astrometric targets.  
Since we used strong maser signals as the phase reference, 
in order to allow the detection of weak background sources close in angle 
on the sky, we needed to measure the maser positions.
Because the 12 GHz methanol masers are often spread over a region
of $\sim1''$, we had to determine their positions from our VLBA data.
We accomplished this in one of two ways.
If one of the background sources had a well determined position
($<10$~mas) and we could image it after phase referencing to the
maser (even with a poorly determined {\it a priori} maser position), 
then we iteratively refined the maser position to a similar accuracy.

When the maser position cannot be measured against a known source
by phase referencing, one can take advantage of the excellent 
correlator model and map of a strong maser spot by simply 
Fourier transforming the data, 
following amplitude calibration and atmospheric model correction, 
but {\it without} phase referencing (\ie after Calibration ``step 3''
describe in \S\ref{calib}).  
This produces a very crude image, which resembles optical speckles, 
owing to the large phase excursions introduced mostly by short-term 
fluctuations in the atmospheric delay. 
For our data, the brighter speckles were concentrated in a
region of about 50 to 100~mas radius, and we were able to estimate the
centroid of the speckle distribution to an accuracy of about 10 to 20~mas.  

Table~\ref{table:positions} lists the absolute positions the maser 
and background continuum sources, as well as the observed brightness 
and restoring beam sizes at the first epoch.

\begin{deluxetable}{lllcrlc}
\tablecolumns{7} \tablewidth{0pc} \tablecaption{Positions and Brightnesses}
\tablehead {
  \colhead{Source} & \colhead{R.A. (J2000)} &  \colhead{Dec. (J2000)} &
  \colhead{$\theta_{\rm sep}$} & \colhead{P.A.} &
  \colhead{$T_b$} & \colhead{\Vlsr} 
\\
  \colhead{}       & \colhead{(h~~m~~s)} &  \colhead{(d~~'~~'')} &
  \colhead{(deg)} & \colhead{(deg)} &
  \colhead{(Jy/b)} & \colhead{(\kms)}
            }
\startdata
 \Ssrc ..............& 06 08 53.3395  &+21 38 29.100 & ... & ... & $6-15$  & 10.8 \\
 J0603+2159 ...   & 06 03 51.55708 &+21 59 37.6982 & 1.2 & $-73$ &0.080  &...     \\
 J0607+2129 ...   & 06 07 59.5657  &+21 29 43.720  & 0.3 &$-125$ &0.005  &...     \\
 J0607+2218 ...   & 06 07 17.4360  &+22 18 19.080  & 0.5 &$-132$ &0.008  &...     \\
 J0608+2229 ...   & 06 08 34.3109  &+22 29 42.981  & 0.2 &$-153$ &0.056  &...     \\
 \\
 \Gsrc     ..... & 07 32 09.7833  &$-$16 58 12.589   & ... & ... &$3-7$  & 22.8   \\
 J0735$-$1735 ...   & 07 35 45.81246 &$-$17 35 48.5027 & 1.1 &126 &0.29&...       \\
 J0729$-$1636 ...   & 07 29 23.93    &$-$16 36 56.2    & 0.8 &$-62$ &... &...     \\
\enddata
\tablecomments {  
  Absolute positions are accurate to about $\pm1$~mas and
  are based on the positions of the ICRF catalog sources 
  J0603+2159 and J0735--1735, whose positions
  are known to about $\pm0.5$~mas in each coordinate.
  Angular offsets ($\theta_{\rm sep}$) and position angles (P.A.) east of north 
  relative to the maser source are indicated in columns 4 and 5. 
  Brightnesses ($T_b$) for the background sources are from the first epoch.
  Restoring beam sizes (FWHM) were 1.0 mas (round) for \Ssrc\ and its
  background calibrators and $8.0\times2.8$ mas$^2$ at a P.A. of $15^\circ$ east of north
  for \Gsrc and its calibrators.
  }
\label{table:positions} 
\end{deluxetable}

\section{Observations and Data Analysis}

\subsection  {Observations}

Our observations were conducted under VLBA program BR100B.
We observed \Ssrc\ and \Gsrc\ over 8-hour tracks at four epochs:
2005 October 13, 2006 April 6 and October 4, and 2007 March 25.
These dates well sample the peaks of the sinusoidal 
trigonometric parallax signature in right ascension.
The declination signature is significantly smaller than that 
for right ascension, and we made no effort to detect it.
This sampling provides maximum sensitivity for parallax detection
and ensures that we can separate the secular proper motion 
(caused by projections of Galactic rotation, as well as any peculiar 
motion of the masers and the Sun) from the sinusoidal parallax effect.
  
For \Ssrc\ we observed four background sources, while for \Gsrc\ we
observed two background sources.   Table 1 lists the positions of
these sources.  We alternated between two $\approx20$~min blocks,
each consisting of observations of a maser target and its background
sources.  Within a block, we switched sources every 40~s, 
typically achieving 30~s of on-source data.  The observing sequence
for the \Ssrc\ block was four repetitions of \Ssrc, J0607+2129, \Ssrc, 
J0607+2218, \Ssrc, J0608+2229, \Ssrc, J0603+2159.  A similar sequence
was used for the \Gsrc\ block; however, with only two background
sources, we used eight repetitions of such a cycle.  We used the methanol 
maser as the phase-reference source, because it is considerably stronger 
than the background sources and could be detected on individual
baselines in the available on-source time.  

We placed observations of two strong sources (J0530+1331 and J0555+3948) 
near the beginning, middle and end of the observations in order to 
monitor delay and electronic phase differences among the IF bands.
In practice, however, we found minimal drifts and used only a single
scan for this calibration.  We also used data from these strong sources
to check the variations in phase across the bandpasses and found
them to vary by $<5^\circ$ across the central 90\% of the band.
Since the masers were observed near band center and we discarded
the outer channels for the continuum sources, we made no bandpass
corrections.

The rapid-switching observations employed four adjacent frequency bands of
4 MHz bandwidth and recorded both right and left circularly polarized
signals.  The four (dual-polarized) bands were centered at 
Local Standard of Rest velocities ($\Vlsr$) of
108.54, 10.00, $-88.54$ and $-187.08$ \kms\ for \Ssrc, and at
120.54, 22.00, $-76.54$ and $-175.08$ \kms\ for \Gsrc.  For both sources,
the masers were contained in the second band.

\subsection{Correlation of Recorded VLBA Data}

The raw data recorded on tape (or more recently on transportable disks)
at each antenna were shipped to the VLBA correlation facility in
Socorro, NM.  The data from individual antenna pairs were cross-correlated 
with an integration time of 0.92~sec.  Integration times were 
kept short to allow position shifting of the data to accommodate a priori
uncertainties in the maser positions.  At this integration time,
we could not process all eight frequency bands in one pass with 
sufficient spectral resolution for the maser band, without 
exceeding the maximum correlator output rate.  Thus, we correlated 
the data in two passes.  One pass was processed with 16 spectral 
channels for each of the eight frequency bands. This data was used for 
the geodetic blocks (to determine atmospheric delays and clock 
drifts) and for the background continuum sources observed in 
rapid-switching (phase-referencing) mode.  Another pass was 
processed with 256 spectral channels, but only for the single 
(dual-polarized) frequency band containing the maser signals,
giving a velocity resolution of 0.38~\kms, assuming a rest frequency of 
12178.597~MHz for the $2_0 \rightarrow 3_{-1}$~E transition of methanol.

\subsection{Calibration} \label{calib}

We calibrated the correlated data using the NRAO Astronomical Image 
Processing System (AIPS).  The calibration sequence included four steps. 
Step 1 involved correction of interferometer delays and phases for
the effects of diurnal feed rotation (parallactic angle),
for errors in the values of the Earth's orientation parameters
used at the time of correlation, and for any small position shifts 
in the a priori coordinates of either the maser or background sources.
Since the VLBA correlator model includes no ionospheric delays, we used 
global total electron content models to remove ionospheric effects.
At this point, we also corrected the data for residual zenith atmospheric 
delays and clock drifts (determined from the geodetic block data).  

\clearpage

\begin{figure}
\epsscale{0.9} 
\plotone{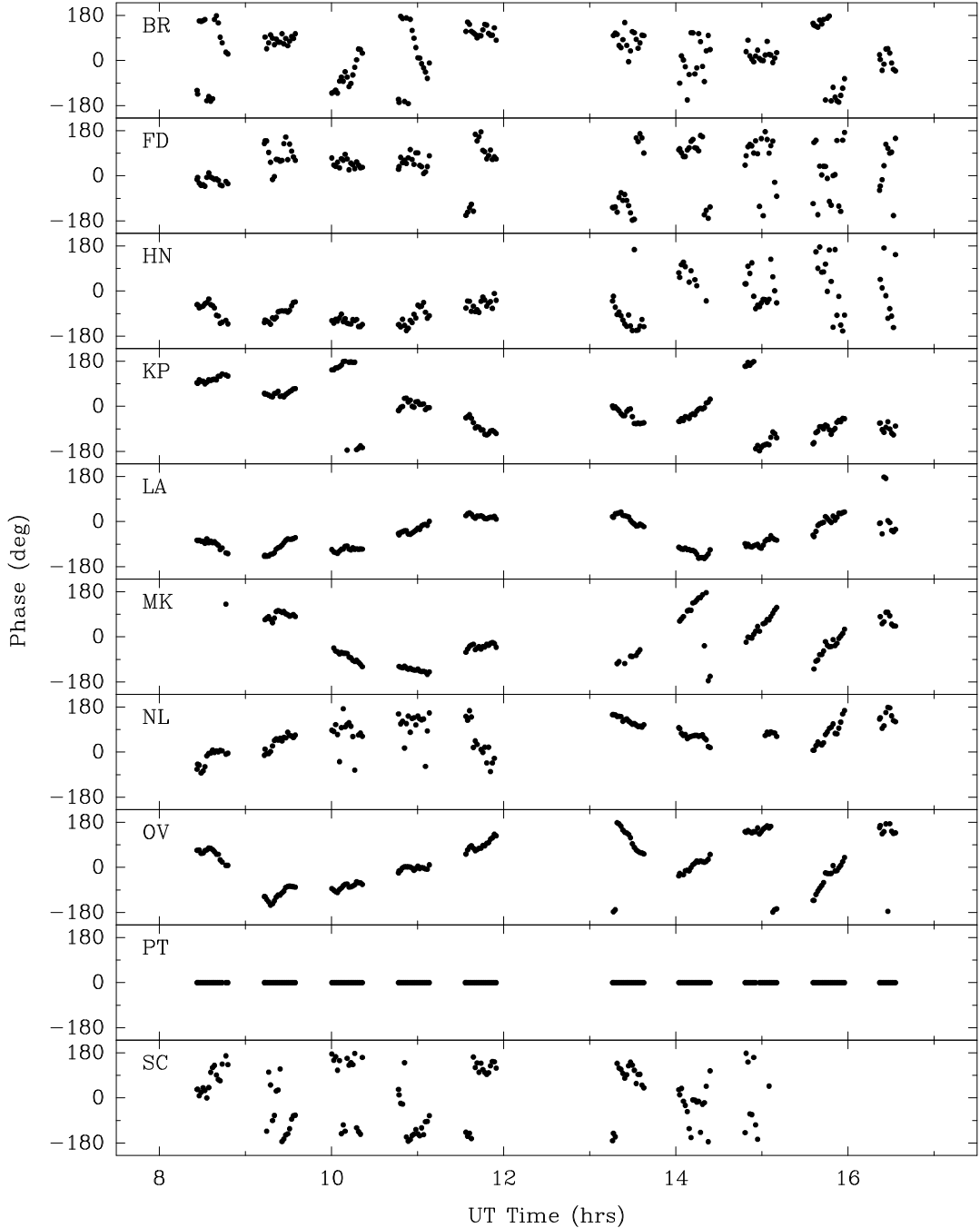} 
\caption{
  Antenna-based phases from maser emission of \Ssrc\ at 
  $\Vlsr = 11.8$~\kms\ used to phase reference both the maser
  and background continuum source interferometer data.
  These data were taken from the first epoch observations on
  2005 Oct 13.
  VLBA antennas are labeled in the upper left corner of each 
  panel.  Interferometer phases are defined between $-180^\circ$ 
  and $+180^\circ$ and wrap across these boundaries.  For most
  antennas at most times, the phase is contained within a few
  cycles, indicating calibration of interferometer elements to
  within a few wavelengths.  Where the baseline phases, constructed
  from these antenna phases, exhibited jumps greater than $60^\circ$,
  we flagged the data between them on a baseline
  by baseline basis.
  \label{s252_phases}
        }
\end{figure}

\clearpage

In step 2 we adjusted interferometer visibility amplitudes for 
the small (few \%) effects of biases in the threshold levels of the data
samplers at each station.  We also entered system temperature and
antenna gain curve information into calibration tables, which allow
conversion of correlation coefficients to flux density units.

In step 3, we performed a ``manual phase-calibration'' to remove
delay and phase differences among all eight 4-MHz bands.
This was done by selecting one scan on a strong calibrator, fitting
fringe patterns to the data for each frequency band, and then shifting 
delays and phases to remove offsets.  At this point, the interferometer
data had delay errors less than $\approx0.03$~nsec ($\approx1$~cm
of excess path) for sources observed at high elevations.  
Finally, we shifted the frequency axes of the maser interferometer spectra
to compensate for the Doppler shift caused by changing projection 
of the Earth's orbit and spin toward the source, in order to keep a 
constant \Vlsr\ in any given spectral channel.

After step 3, the visibility phases are calibrated for all known
effects and residual phases are typically dominated by 
short-term atmospheric variations of $\sim3$~cm and nearly
constant absolute position errors usually less than $\sim3$~mas
(see \S\ref{abs_pos} and Table~\ref{table:positions}).  
Atmospheric variations cancel, to first order in the source separations 
(typically $1^\circ$ or 0.02 radians on the sky), when constructing 
phase differences (``phase-referencing'') between the target maser 
and the background continuum source.   
Absolute position errors are largely constant over our observations 
and cancel over time when fitting for a parallax.
(Only the time-variable position changes, owing to parallax and
proper motion, will not cancel in time.  Typical Galactic proper motions
of $\approx1$~\masy\ result in differential position shifts of
typically 0.02~mas, when the galactic source (\eg maser) is used
as the phase reference.)  

The final calibration, step 4, involved selecting a maser phase-reference.
We used a single spectral channel of the maser source, selected
to be strong on the longest interferometer baselines, and hence
compact.  The reference phases from spectral channel 128 at 
$\Vlsr = 10.8$~\kms\ of \Ssrc\ are shown in Fig.~\ref{s252_phases}.
The flux density of this maser was typically $\sim10$ Jy.
For most antennas at most times the phases were easily ``connected''
and interpolated to the time of the background source observations.
(There was one background source observation between each pair of
\Ssrc\ phase-reference observations.)   When the differences
between adjacent reference phases exceeded $60^\circ$,  
the data between those times was discarded.  
The limit of $60^\circ$ was chosen to be large enough to allow
for (calibratable) short-term atmospheric fluctuations, but not
too large so as to encounter turn ambiguities when interpolating
between reference phases.  Note that this editing
was done on baseline (not antenna) phases, since baseline phases are
what are ultimately used when calibrating data.

\begin{figure}
\epsscale{0.8} 
\plotone{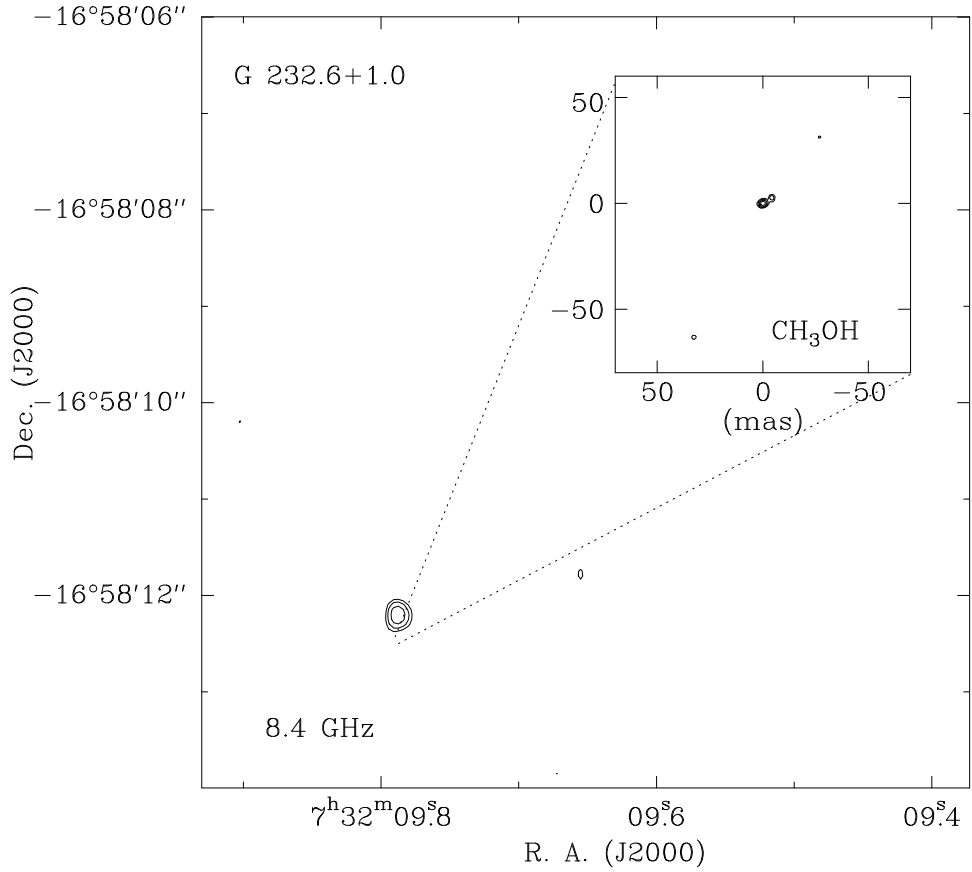} 
\caption{
  Image of continuum emission associated with a hyper-compact 
  HII region in \Gsrc, generated from archival VLA data, with
  the velocity-integrated maser emission inset.  The reference
  maser spot is at the origin of the inset and was the only
  spot used for the parallax fits.  Contour levels start at
  0.55 mJy and increase by factors of 1.4 for the HII region, and 
  they start at 
  0.33~Jy~\kms\ and increase by factors of 2 for the maser emission.  
  Zero contours are suppressed for clarity. 
  \label{g232_overlay}
        }
\end{figure}

For \Gsrc, the sole strong maser feature presented as a 
asymmetric double with the strongest component having a flux density
of $\sim5$~Jy (see inset in Fig.~\ref{g232_overlay}).  
Using this spectral channel as a phase reference 
initially resulted in the appearance of a false double-structure in the
background continuum source image.   To remedy this problem,
we performed an iterative self-calibration sequence on the maser channel
in order to construct a good image of that maser double.  We then
used this image to remove the effects of the maser structure when
calculating reference phase tables.

\section{Parallax and Proper Motion Fitting}

After calibration, we Fourier transformed the gridded \uvdata, using
the AIPS task IMAGR, to make images of the maser emission in all spectral 
channels and the background continuum sources.  The images were 
deconvolved with the point-source response using the CLEAN algorithm.
In order to provide the data needed to measure parallax and proper 
motions, we fitted an elliptical Gaussian brightness distribution 
to the brightest maser spots and the background continuum sources at 
all four epochs, using the task JMFIT.
In general, our background sources are compact and appear 
dominated by a single component.  The (generally weak) extragalactic 
continuum sources we find usually are very compact and have little
internal structure that could complicate astrometric accuracy.
This is in contrast to well known quasars with bright jets, such as
superluminal sources, for which structural changes would be problematic. 

The change in position of the maser spot(s) relative to the background 
continuum source(s) was then modeled by the parallax
sinusoid in both coordinates, completely determined by one parameter
(the parallax) and a secular proper motion in each coordinate.
The model included the effects of the ellipticity of the Earth's orbit.

\clearpage

\begin{figure}
\epsscale{0.8} 
\plotone{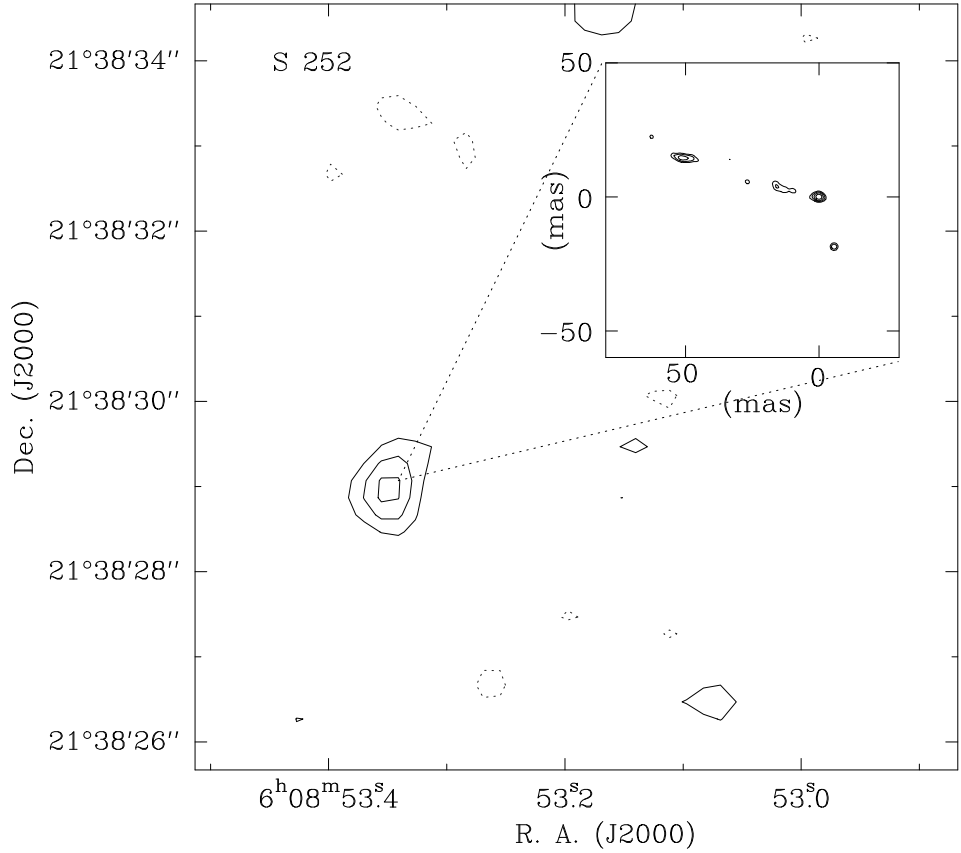} 
\caption{
  Image of continuum emission associated with a hyper-compact 
  HII region in \Ssrc, generated from archival VLA data, with
  the velocity-integrated maser emission inset.  The reference
  maser spot is at the origin of the inset. Maser spots used
  for the parallax fits are at $(0,0)$ and $(-5,-18)$~mas.  Contour levels are
  linearly spaced at 0.08~mJy for the HII region, and they start
  at 3~Jy~\kms\ and increase by factors of 2 for the maser emission.  
  Zero contours are suppressed for clarity. 
  \label{s252_overlay}
        }
\end{figure}

\clearpage

The choice of data and their weighting in the parallax and proper motion 
fitting is complicated by two issues.  Firstly, some maser spots
are blended in position and frequency and the intensity of the blended
components can vary with time.  Generally, we fitted positions in all
spectral channels where the maser spots looked to be good candidates for 
astrometry (\ie\ compact emission, minimal blending, etc.).  We then
fitted a parallax and proper motion to each spot separately in order
to assess its astrometric quality, based on the magnitude of the post-fit
residuals.  Usually only a small number of maser spots were passed along
for final fitting.  Secondly, the formal position uncertainties 
were often unrealistically small, since (a priori unknown) sources 
of systematic error often dominate over random noise.

The north-south components of relative positions often
have greater uncertainty than the east-west components, because
(1) interferometer beams are generally larger in the north-south direction and 
(2) systematic errors from unmodeled atmospheric delays usually
are more strongly correlated with north-south positions \citep{Honma:08}.  
In order to allow for such systematic errors, we assigned independent 
``error floors'' to the east-west and north-south position data and 
added these floors in quadrature with the formal position-fitting 
uncertainties.  Trial parallax and proper motion fits were conducted and 
a separate reduced $\chi^2$ (per degree of freedom) statistic was 
calculated for the east-west and north-south residuals.  The error floors 
were then adjusted iteratively so as to make the reduced 
$\chi^2\approx1.0$ for each coordinate.

Individual maser spots were allowed to have different proper motions,
owing to internal motions that are typically a few~\kms\ for 12-GHz 
methanol masers.  However, since we expect no detectable proper
motion for the extragalactic sources, in the cases where we had more
than one extragalactic source, we constrained the proper motion of
any given maser spot to be the same when measured against all 
extragalactic sources.  Note that we used a single east-west and
north-south error floor in any least-square fit and these floors
are the same for the combined and the individual fits.  The small
differences in error floors among the fits can result in the combined
fit parameters differing slightly from a simple combination of the
individual fit values.

\subsection   {\Ssrc}

\clearpage

\begin{figure}
\epsscale{0.8} 
\plotone{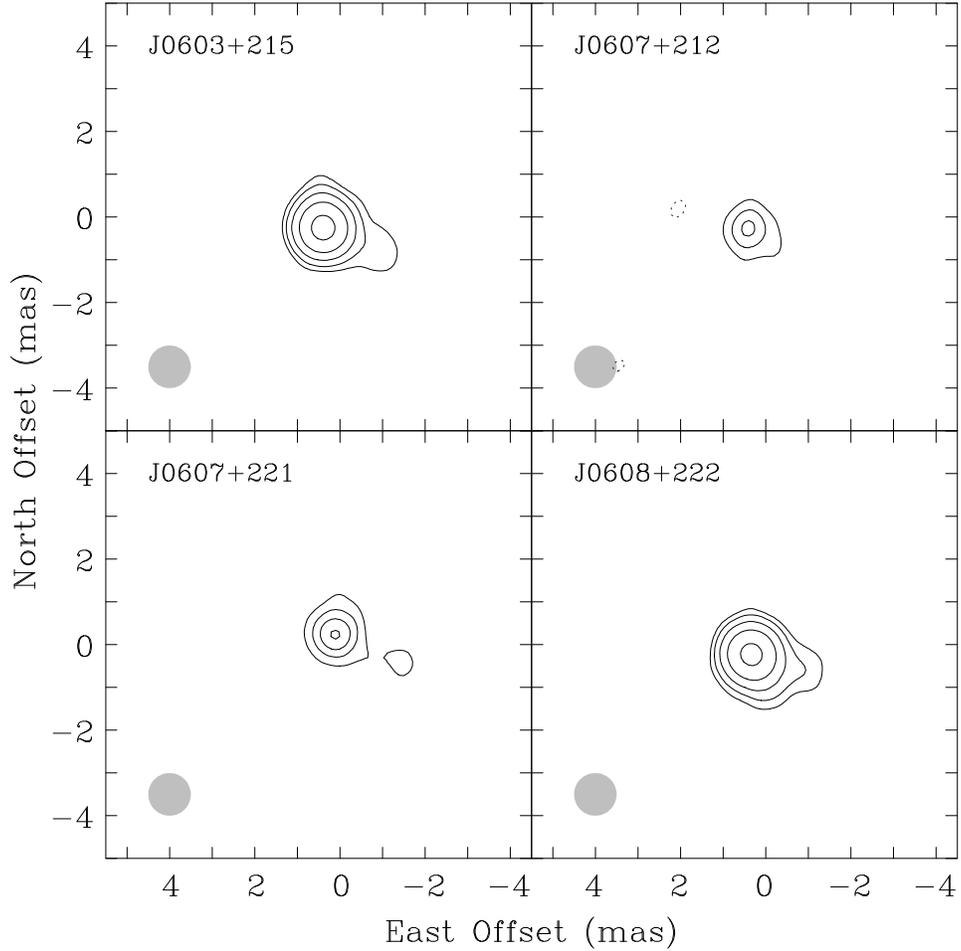} 
\caption{
  Images of background continuum sources near \Ssrc.
  Source names are in the upper left
  corner and restoring beams are in the lower left corner
  of each panel.  All images are from the first epoch observations
  on 2005 Oct 13.  For the stronger sources, contour levels
  are spaced by factors of 2, starting at 4~mJy for J0603+2159
  and 3~mJy for J0608+2229.  For the weaker sources, contour
  levels are spaced linearly at 1.5~mJy increments for J0607+2129
  and 2~mJy for J0607+2218. Zero contours are suppressed for clarity.
  The weakest source, J0607+2218, was not used for the parallax  
  measurement of \Ssrc.
  \label{s252_qsos}
        }
\end{figure}

\clearpage

For the \Ssrc\ sources, the 
interferometer (``dirty'') beam had a FWHM of 1.3 by 0.6 mas 
at a PA of $-5^\circ$ East of North.  In the deconvolution 
(``CLEAN'') step, we used a round restoring beam of 1.0~mas 
in order to simplify the fitting of potentially close maser spots. 

\clearpage 

\begin{figure}
\epsscale{1.0}
\plotone{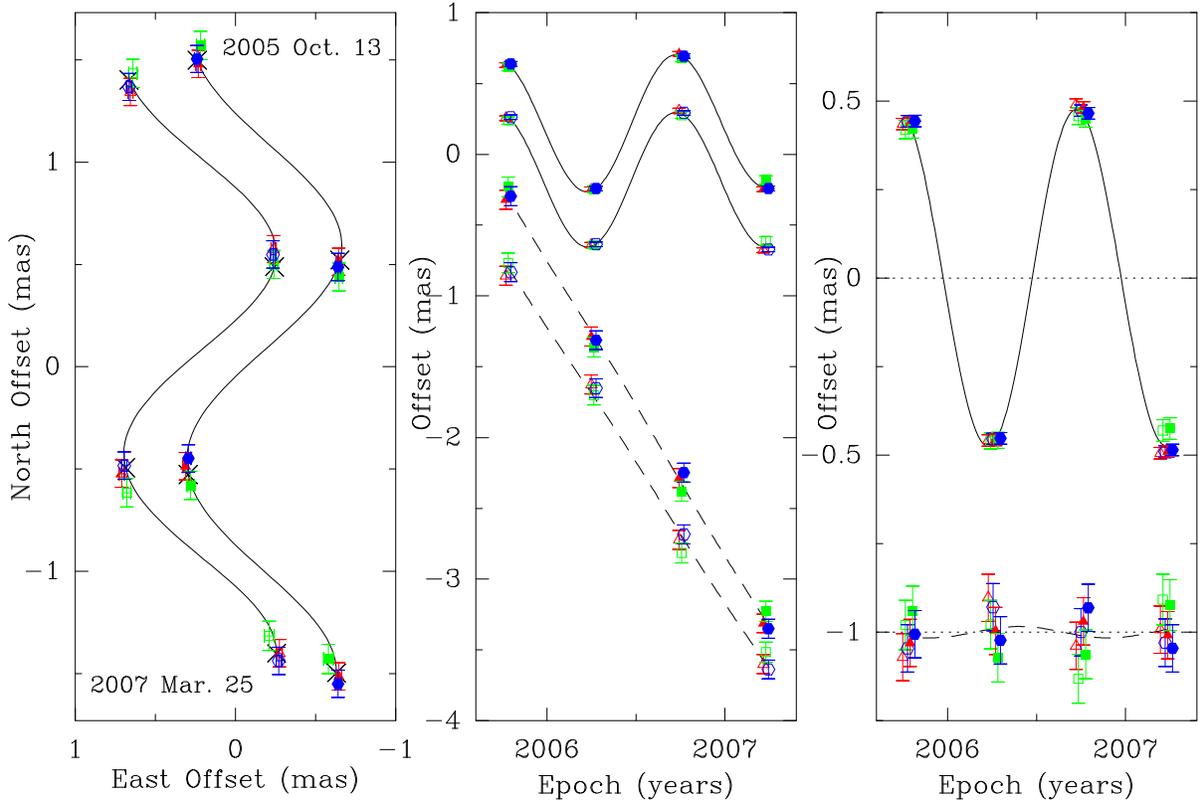} 
\caption{
  Parallax and proper motion data and fits for \Ssrc.
  Plotted are position measurements of two maser spots {\it (open and solid symbols)} 
  relative to the three background sources: J0603+2159 {\it (red triangles)}, 
  J0607+2218 {\it (green squares)} and J0608+2229 {\it (blue hexagons)}.
  {\it Left Panel:} Positions on the sky with first and last epochs labeled.
  Data for the two maser spots are offset horizontally for clarity.
  The expected positions from the parallax and proper motion fit
  are indicated {\it (crosses)}.
  {\it Middle Panel:} East {\it (solid lines)} and North {\it (dashed lines)} position
  offsets and best fit parallax and proper motions fits versus time.  
  Data for the two maser spots are offset vertically, the northward
  data have been offset from the eastward data, 
  and small time shifts have been added to the data for clarity
  {\it Right Panel:} Same as the {\it middle panel}, except the
  with the best fit proper motions have been removed, allowing
  all data to be overlaid and the effects of only the parallax seen.
  \label{s252_fits}
        }
\end{figure}

\clearpage

\begin{deluxetable}{lllll}
\tablecolumns{5} \tablewidth{0pc} 
\tablecaption{\Ssrc\ Parallax \& Proper Motion Fits}
\tablehead {
  \colhead{Maser \Vlsr} & \colhead{Background} &  
  \colhead{Parallax} & \colhead{$\mu_x$} &
  \colhead{$\mu_y$} 
\\
  \colhead{(\kms)}      & \colhead{Source} & 
  \colhead{(mas)} & \colhead{(\masy)} &
  \colhead{(\masy)}
            }
\startdata
 10.8 ......& J0603+2159 &$0.485\pm0.016$ &$+0.00\pm0.03$ &$-1.94\pm0.10$ \\
 10.8 ......& J0607+2218 &$0.462\pm0.012$ &$+0.03\pm0.02$ &$-1.95\pm0.14$ \\
 10.8 ......& J0608+2229 &$0.477\pm0.009$ &$-0.01\pm0.02$ &$-1.96\pm0.07$ \\
 11.5 ......& J0603+2159 &$0.481\pm0.013$ &$+0.03\pm0.02$ &$-2.07\pm0.04$ \\
 11.5 ......& J0607+2218 &$0.459\pm0.012$ &$+0.06\pm0.02$ &$-2.08\pm0.13$ \\
 11.5 ......& J0608+2229 &$0.473\pm0.009$ &$+0.02\pm0.01$ &$-2.10\pm0.09$ \\
\\
 10.8 ......& combined  &$0.476\pm0.006$ &$+0.00\pm0.01$ &$-1.95\pm0.04$ \\
 11.5 ......&           &                &$+0.03\pm0.01$ &$-2.08\pm0.04$ \\
\enddata
\tablecomments {Combined fit used a single parallax parameter
for both maser spots relative to the three background sources;
a single proper motion was fit for each maser spot relative 
to all three background sources. The quoted uncertainty for the combined
parallax fit is the formal fitting uncertainty multiplied by $\sqrt{2}$
to allow for the possibility of correlated positions for the 2 maser spots.} 
\label{table:s252_ppm}
\end{deluxetable}

In Fig.~\ref{s252_overlay}, we show VLA archive data of 8.4 GHz 
continuum emission from an associated hyper-compact HII region, with
the first epoch image of the maser emission integrated over all 
spectral channels with detectable emission. 
In Fig.~\ref{s252_qsos}, we show an image at the first epoch of 
the four background radio sources.  
The small departures from 
a point-like image at the lowest contour levels is probably a
result of small imperfections in the phase calibration transfered 
from the maser data.

The position data from three of the four 
background continuum sources produced consistent results with small
residuals.  The data for J0607+2129, our weakest source, yielded 
significantly larger post-fit residual  and was not used for the final 
parallax and proper motion fitting.  The fitting results are presented 
in Table~\ref{table:s252_ppm}.  Our measured parallax for \Ssrc\ is 
$0.476 \pm 0.006$~mas; the quoted uncertainty is the formal error
multiplied by $\sqrt{2}$ to allow for the possibility of correlated
position variations for the two maser spots.  This could result
from small variations in the background source or from
unmodeled atmospheric delays, both of which would affect the two maser
spots nearly identically.  

Our parallax for \Ssrc\ corresponds to a distance of 
$2.10^{+0.027}_{-0.026}$~kpc.  The average proper motion of the two 
astrometric maser spots is $0.02\pm0.01$~\masy\ toward the East
and $-2.02\pm0.04$~\masy\ toward the North.  At the distance
implied by the parallax measurement, these motions correspond
to $0.3$~\kms\ and $-20.1$~\kms\ eastward and northward, 
respectively.  Completing the full space velocity, the average LSR 
velocity of the spots is $11.2$~\kms, which corresponds to a 
heliocentric radial velocity of $23.6$~\kms.  The space motion of 
the two masers spots are almost identical, and, since internal
motions of methanol masers are typically small
\citep{Moscadelli:02}, we expect the maser space motions to follow
that of the central young stellar object to within about $\pm3$ \kms.

\subsection {\Gsrc}

\clearpage

\clearpage

For \Gsrc, the reference maser spot was considerably larger than in \Ssrc, 
and we could only effectively use the inner five VLBA stations for 
imaging.  The dirty beam was about 8 by 3~mas at a PA of $15^\circ$.  
Because of the limited interferometer data for this low Declination 
source, we used a restoring beam equal to the FWHM of the dirty beam.

\begin{figure}
\epsscale{0.45} 
\plotone{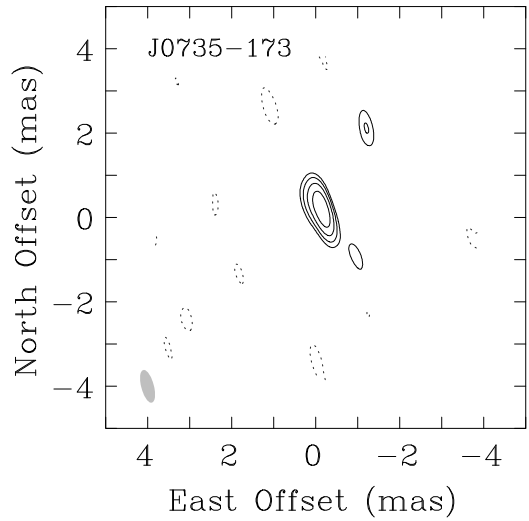} 
\caption{
  Image of background continuum source, J0735-1735, used for the parallax
  measurements of \Gsrc.  The image is from the first epoch observations
  on 2005 Oct 13.   The restoring beam ({\it gray}) is in the lower left 
  corner.  Contour levels are spaced by factors of 2, starting at 
  20~mJy.  The zero contour is suppressed for clarity. 
  \label{g232_qso}
        }
\end{figure}

\begin{figure}
\epsscale{1.0} 
\plotone{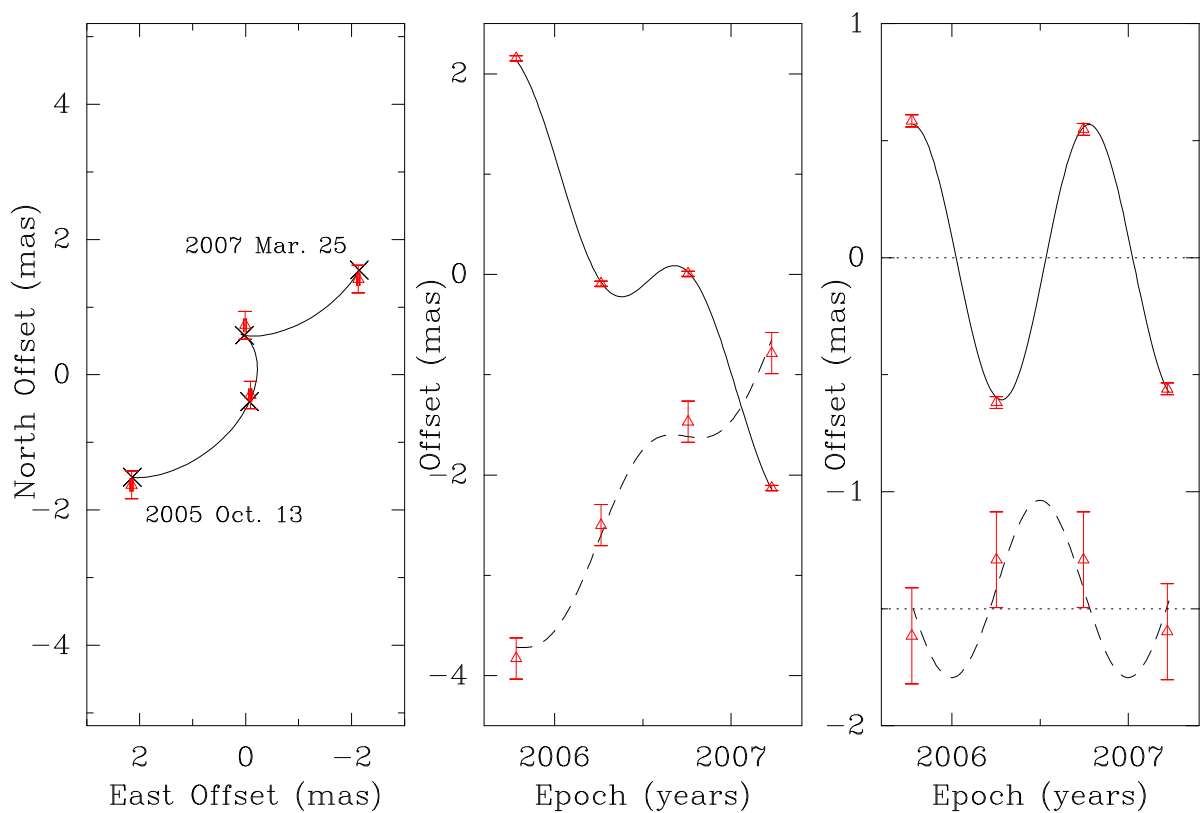} 
\caption{
  Parallax and proper motion data and fits for \Gsrc.
  Plotted are position measurements of the most compact maser spot 
  relative to the background source J0735-1735 {\it (red triangles)}. 
  {\it Left Panel:} Positions on the sky with first and last epochs labeled.
  The expected positions from the parallax and proper motion fit
  are indicated {\it (crosses)}.
  {\it Middle Panel:} East {\it (solid lines)} and North {\it (dashed lines)} position 
  offsets and best fit parallax and proper motions fit versus time.
  The northward data have been offset from the eastward data for clarity.  
  {\it Right Panel:} Same as the {\it middle panel}, except the
  with the best fit proper motion has been removed, allowing
  the effects of only the parallax seen.
  \label{g232_fits}
        }
\end{figure}

In Fig.~\ref{g232_qso}, we show an image from the first epoch of 
of the background radio source J0735-1735.  The second background
source observed, J0729-1636, was not detected.
We plot the position of the \Vlsr=22.8~\kms\ maser spot relative to 
J0735-1735 in Fig.~\ref{g232_fits}. 

\clearpage

\clearpage

Our measured parallax for \Gsrc\ is $0.596\pm0.035$~mas, which
corresponds to a distance of $1.68^{+0.11}_{-0.09}$~kpc.  The proper motion
of the single maser spot is $-2.17\pm0.06$ toward the East and
$2.09\pm0.46$~mas toward the North.  For the measured parallax,
the proper motions correspond to $-17$~\kms\ and $+17$~\kms\ 
eastward and northward, respectively.  

Unlike for \Ssrc, where we had multiple spots and background sources, 
for \Gsrc\ we found only one maser and one background source suitable 
for astrometric measurements. 
The VLBA calibrator survey shows that in 2002 at 8.6 GHz the background 
source, J0735--1735, had a second component of $\approx0.05$~Jy beam$^{-1}$
separated by $\approx4$~mas at a position angle of $\approx40^\circ$
east of north from the strongest component of $\approx0.20$~Jy beam$^{-1}.$
At 12.2~GHz we found J0735--1735 to be dominated by a single component
with peak brightnesses of 0.29, 0.24, 0.31 and 0.32~Jy beam$^{-1}$ at
our four epochs.
The apparent variations in brightness are small and consistent with
a nearly constant brightness source that is imaged in the presence
of typical calibration errors, especially variable phase-referencing
coherence losses at different epochs.
 
In light of the potential limitations of a single background source
and a single maser spot, and because there was effectively only one 
degree of freedom in the fit, which is dominated by the east-west data, 
we have doubled the formal uncertainties as a precaution.
Table~\ref{table:g232_ppm} summarizes the parameters of the parallax 
and proper motion fits.

\clearpage

\clearpage

\begin{deluxetable}{lllll}
\tablecolumns{5} \tablewidth{0pc} 
\tablecaption{\Gsrc\ Parallax \& Proper Motion Fit}
\tablehead {
  \colhead{Maser \Vlsr} & \colhead{Background} &  
  \colhead{Parallax} & \colhead{$\mu_x$} &
  \colhead{$\mu_y$} 
\\
  \colhead{(\kms)}      & \colhead{Source} & 
  \colhead{(mas)} & \colhead{(\masy)} &
  \colhead{(\masy)}
            }
\startdata
 22.8 ......&J0735--1735 &$0.596\pm0.035$ &$-2.17\pm0.06$ &$+2.09\pm0.46$\\
\enddata
\tablecomments {We conservatively quote uncertainties that are double the 
formal fitting values, since effectively there is only one degree of freedom 
for the parallax and proper motion solution which is dominated by the 
east-west data.}
\label{table:g232_ppm}
\end{deluxetable}

\section{Galactic 3-D Motions}

In order to study the 3-dimensional motion of the maser sources in the Galaxy, 
we need to convert their radial and proper motions from the equatorial
heliocentric reference frame in which they are measured into a
Galactic reference frame.  A convenient Galactic frame is one
rotating with a circular velocity $\Theta(R)$ at the position of the source: 
ie, a ``local standard of rest'' at the location of maser.  We will describe this 
procedure in detail in \citet{Reid:09}.

Conversion to a rotating Galactic reference frame depends upon
several parameter values.  The IAU standard value for the distance 
of the Sun to the Galactic center is $\Ro= 8.5$ kpc 
and for the rotation speed of the LSR is $\To = 220$ \kms.  
We assume a flat rotation curve, \ie\ $\Theta(R) = \To$.  
(The effects of a non-flat rotation curve will be considered Paper~VI.) 
Finally, we use the Solar Motion values 
$U_\odot=10.0\pm0.4$,  $V_\odot=5.2\pm0.6$ and $W_\odot=7.2\pm0.4$ \kms, 
derived from Hipparcos data by \citet{Dehnen:98}.   
For these parameter values, our parallax, \Vlsr\ and proper motion 
results for \Ssrc\ imply ``peculiar motions'' (relative to circular motion) 
of $-4\pm3$~\kms\ toward the Galactic center,
$-16\pm1$~\kms\ in the direction of Galactic rotation, and
$-2\pm1$~\kms\ toward the North Galactic Pole.  
For \Gsrc\ we find peculiar motion components of 
of $-4\pm3$~\kms\ toward the Galactic center,
$-10\pm3$~\kms\ in the direction of Galactic rotation, and
$0\pm2$~\kms\ toward the North Galactic Pole.  Thus, both sources
are orbiting the Galaxy slower than it spins (\ie\ slower than
circular rotation).  

We note that evidence from proper motion of Sgr~A*, 
the super-massive black hole at the center of the Galaxy, strongly 
suggests that $\To/\Ro=29.5$~\kms~kpc$^{-1}$, with an uncertainty
less than 1\% \citep{Reid:04}.  Thus,
were we to adopt $\Ro = 8.0$ kpc \citep{Reid:93},
this would require $\To=236$ \kms.  For these Galactic parameters,  
the peculiar velocity components change only slightly:
for \Ssrc\ they become $-3$~\kms\ toward the Galactic center,
$-16$~\kms\ in the direction of Galactic rotation, and
$-2$~\kms\ toward the North Galactic Pole, and
for \Gsrc\ they become $0$~\kms\ toward the Galactic center,
$-10$~\kms\ in the direction of Galactic rotation, and
$0$~\kms\ toward the North Galactic Pole.   Thus, changing 
\Ro\ and \To\ by amounts consistent with current levels of 
uncertainty would not change the conclusion that these star forming
regions orbit the Galaxy slower than it spins.

Research on the structure of the Galaxy in Nanjing
University is supported by the National Science Foundation of
China under grants 10133020, 10103003 and 10373025, and 
A. Brunthaler is supported by DFG Priority Programme 1177.

\end{document}